# Low-threshold Stimulated Emission of Hybrid Perovskites at Room Temperature through Defect-Mediated Bound Excitons

Jiangjian Shi,[1] Yiming Li,[1,2] Jionghua Wu,[1,2] Huijue Wu,[1] Yanhong Luo,[1,2] Dongmei Li,[1,2] Jacek J. Jasieniak,[3,*] Qingbo Meng[1,4,*]

1. Key Laboratory for Renewable Energy, Chinese Academy of Sciences, Beijing Key Laboratory for New Energy Materials and Devices, Institute of Physics, Chinese Academy of Sciences, Beijing 100190, P. R. China.

2. Department of Physics Science, University of Chinese Academy of Sciences, Beijing 100049, P. R. China

3. ARC Centre of Excellence in Exciton Science, Department of Materials Science and Engineering, Monash University, Clayton, Victoria 3800, Australia.

4. Center of Materials Science and Optoelectronics Engineering, University of Chinese Academy of Sciences, Beijing 100049, China

*Corresponding authors: jacek.jasieniak@monash.edu, qbmeng@iphy.ac.cn

**Abstract:** Excitonic stimulated emission provides a promising mechanism and route to achieve low-threshold semiconductor lasers for micro-nano optoelectronic integrations. However, excitonic stimulated emission from quantum structure-free semiconductors has rarely been realised at room temperature due to the phase transition between excitonic and electron-hole plasma states. Herein, we show that through trap-state and band-edge engineering, bound exciton states can be stabilised within the hybrid lead bromide perovskite. Under modest pumping conditions, these states enable stimulated emission behaviour that exhibits a low threshold carrier density of only $1.6\times10^{17}$ cm$^{-3}$, as well as a high peak gain coefficient of ~1300 cm$^{-1}$. This is the first time that bound exciton stimulated emission has been realised at room temperature from a quantum structure-free semiconductor. Not only does this open up new research horizons for perovskite materials, but also it has important implications for semiconductor excitonic physics and the development of next-generation optoelectronic applications.

Since the discovery of the ruby laser in 1960, lasers have literally changed the world by enabling global communication through satellites and the internet, advancing manufacturing, and unravelling the wonders of almost all the scientific worlds.[1-2] As optoelectronic components continue to reduce in size and increase in speed, the need for integrated microscale laser systems that operate at room temperature will follow.[3-5] Semiconductor thin-film lasers are a promising photonic source for such micro/nano optoelectronic applications, owing to their tunable wavelength, compact size and ease of optical or electronic coupling.[6-8] For satisfying on-chip integration, realising laser devices with low energy consumption is critical. Numerous materials, device structures and physical mechanisms have been investigated to achieve this goal, mainly targeting reduced lasing threshold and increased optical gain.[9-17]

Lasers operate by amplifying stimulated emission (STE) arising from an optically or electrically pumped semiconductor gain medium that is under population inversion.[18] Compared to the conventional degenerate electron-hole plasma (EHP) mechanism, the nature of exciton states and their photon emission processes provide more possibilities to realize population inversion, which can be exploited to significantly reduce the STE thresholds, especially at low temperatures.[18] However, except for in quantum structures,[11-13] room-temperature (RT) STE of exciton states within inorganic semiconductors has rarely been realised.[18] This is attributed to the low Rydberg energy ($R_y^*$) of such materials, for which (i) the thermal energy at RT readily overcomes the necessary attractive exciton Coulomb interactions and (ii) the high carrier pumping required for STE usually induces the excitonic-EHP phase transition.[18] Owing to the high energy bandgap ($E_g$) and $R_y^*$ (~60 meV), ZnO is a standout case that exhibits violet STE at RT with a high threshold of >$10^{18}$ cm$^{-3}$.[10] However, the high $E_g$ of ZnO makes it difficult to be optically or electrically pumped and controversy still exists on whether its STE arises from an excitonic process.[18-19] As such, developing new excitonic material systems that can achieve low-threshold RT STE across the visible spectrum from bulk semiconductor continues to be a major challenge.

Hybrid lead halide perovskites have emerged as lucrative optoelectronic materials for light-emitting and photovoltaic applications.[6,20] They have also been used as optically pumped lasers, exploiting their EHP states to achieve modest STE thresholds of ~$10^{18}$ cm$^{-3}$.[8, 21-26] The population inversion of perovskite EHP originates from the carrier occupation of continuum bands. Achieving EHP within perovskites is a double-edged sword. On the one hand, the high effective density-of-states close to the band-edge of perovskite (~$10^{18}$ cm$^{-3}$) contribute to its high light absorption coefficient ($10^4 \sim 10^5$ cm$^{-1}$); however, they also necessitate large pumping thresholds of >$10^{18}$ cm$^{-3}$ for realising a population inversion.[21] Exploiting the excitonic characteristics of hybrid perovskites presents a major opportunity towards reducing these thresholds.

In this article, we describe how bound exciton complex (BE) states that form by the Coulomb binding between an exciton and a trap centre can be accessed within solution-processed CH$_3$NH$_3$PbBr$_3$ (MAPbBr$_3$) polycrystalline films through precise engineering of the perovskite crystallization, traps and exciton state dynamics. It is shown that these BE states enable RT STE with a nearly 10-fold reduction in threshold fluence compared to previous reports harnessing EHP inversion. This is the first work to report BE STE at RT from a cavity-free bulk semiconductor. Moreover, it presents a lucrative excitonic mechanism and route for optoelectronic applications, helping to broaden the research and application horizons of perovskite materials.

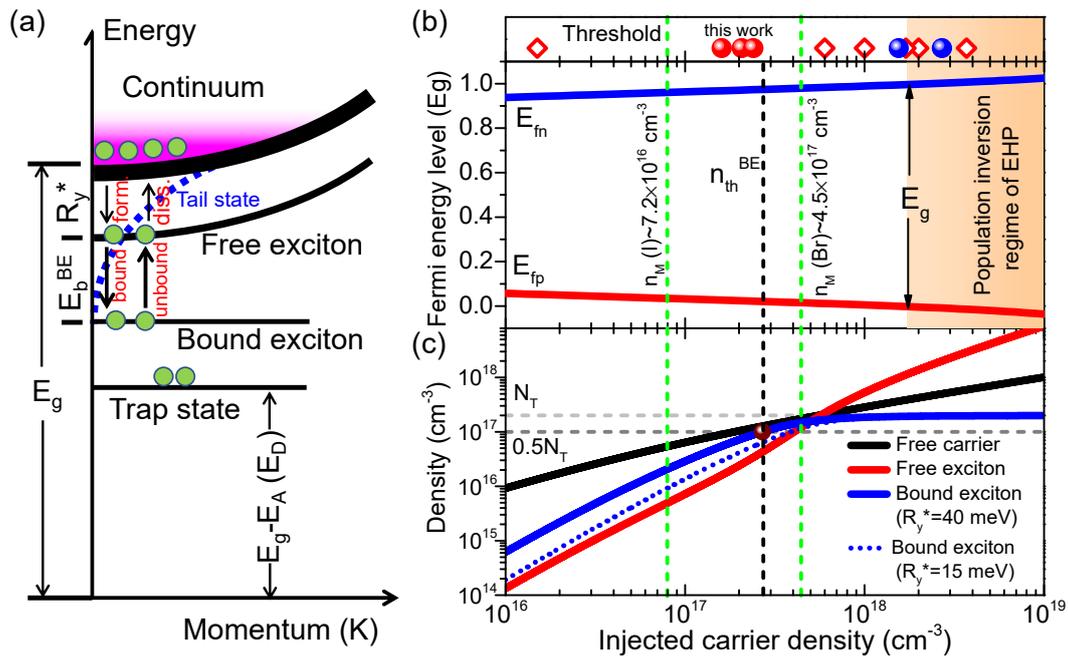

**Figure 1. Charge occupation and population inversion within different energy states of the perovskite.** (a) Schematic diagram of energy states (including continuum band, free and bound excitons, and trap state) and internal charge transfer processes within a perovskite semiconductor. (b) Theoretical calculations of the quasi-Fermi energy level (electron: $E_{fn}$, hole: $E_{fp}$) of the perovskite under carrier injections. The EHP shows a population inversion when the $\Delta\mu$ (i.e. $E_{fn}$-$E_{fp}$) is larger than the energy bandgap ($E_g$), with a threshold of

~1.7×10$^{18}$ cm$^{-3}$. (c) Theoretical calculation for the densities of free carriers, free excitons and bound excitons as determined through the Saha-Langmuir equation and the thermal equilibrium between free and bound excitons. The density of trap centers ($N_T$) that could bind free excitons was experimentally determined to be ~2×10$^{17}$ cm$^{-3}$. Moreover, experimental values of the $R_y^*$ at ~40 meV and the binding energy of an exciton bound to a trap center ($E_b^{BE}$) at ~100 meV were also used. When the occupation probability of trap centers is higher than 1/2, a population inversion of the BE state will occur. The inversion threshold ($n_{th}^{BE}$) is determined to be ~2.7×10$^{17}$ cm$^{-3}$. The Mott densities ($n_M$) of iodide and bromide perovskites are illustrated through dashed green lines crossing (b) and (c). For comparison, the STE thresholds of perovskites reported in the literate (open red squares)[7,8,22-26] and those obtained here (solid circles) are included in (b). The BE STE is depicted in red, while EHP in blue.

**Theoretical estimation of perovskite STE: EHP vs. BE**

While STE has been confirmed in perovskites, its mechanism has not been clearly understood, with most reported cases showing high thresholds of >10$^{18}$ cm$^{-3}$.[7-8, 20-28] As such, compared to conventional semiconductors, the advantages of these material system in terms of STE performance have not been fully realised. Towards addressing this, we firstly give a theoretical description of the energy states and charge-carrier properties in perovskites, and then estimate their population inversion limitations for EHP and excitonic processes. As depicted in Figure 1(a), charge-carriers generated within a semiconductor from photoexcitation or electric injection populate within a variety of energy states, including the continuum band (i.e. conduction and valence bands) and its tail states, free exciton (FE), BE and trap states.[18] The balance between the continuum band carriers and excitons is described by the Saha-Langmuir equation.[29] Meanwhile, the equilibrium between the FE and BE states can be considered analogous to that of detrapped and trapped charges at trap centres, respectively (Supplementary Note 1).[30]

The quasi-Fermi energy level splitting ($\Delta\mu$) of MAPbBr$_3$ is calculated from the Fermi-Dirac distribution to estimate the population inversion of EHP (Figure 1(b)). Under higher carrier density, $\Delta\mu$ becomes progressively larger. When it exceeds the bandgap energy of MAPbBr$_3$ ($E_g$~2.4 eV), a population inversion is achieved. STE can be realised under these conditions within a suitable device configuration. The threshold of the EHP STE for MAPbBr$_3$ is found to be ~1.7×10$^{18}$ cm$^{-3}$, which is similar to other theoretical predictions for MAPbI$_3$.[21] As is shown in the top graph of Figure 1(b), this value is comparable to both previously reported STE values and our experiments without energy-state engineering (i.e. ~$10^{18}$ cm$^{-3}$).[8,23-26] Notably, the EHP mechanism underlying these experimental STE observations can be readily identified from pump fluence-dependent emission characteristics, which show a progressive shift of the STE peak due to bandgap renormalization (red-shift) or carrier filling of continuum bands (blue-shift).[7,18,22,23,28]

To evaluate the population inversion behaviour within excitonic systems, we have determined the occupancy density within the continuum band, as well as FE and BE states, under thermal equilibrium (Figure 1(c), Supplementary Note 1). In these systems, when excitons occupy more than half of the trap centres, a population inversion of the BE state can occur under.[11] In our previous work we proved the existence of the BE state within MAPbBr$_3$.[31] Additionally, we derived experimental binding energies of the FE ($R_y^* \sim 40$ meV) and BE ($E_b^{BE} \sim 100$ meV), as well as the trap densities responsible for the BE (~$2\times10^{17}$ cm$^{-3}$). Using these values, we can determine the threshold for BE STE at ~$2.7\times10^{17}$ cm$^{-3}$. This value is more than 6 times lower than that for STE arising from EHP. It confirms that excitonic processes are more lucrative towards achieving STE than EHP.[18] Moreover, this threshold value of MAPbBr$_3$ is lower than its theoretically calculated Mott density (~$4.5\times10^{17}$ cm$^{-3}$),[18] while the opposite is found for MAPbI$_3$ ($n_M \sim 7.2\times10^{16}$ cm$^{-3}$). Noting that the Mott density provides an estimation of the upper limit for the stable regime of the excitonic phase, these values confirm that the bromide perovskite system is a promising and uncommon candidate to realise low-threshold BE STE.

**Stabilisation of perovskite exciton states**

Until recently, BEs within bulk perovskites have not been widely observed or studied.[31] This has arisen because these exciton states can be easily screened by local electric fields stemming from shallow defect (ST) or tail state induced band-edge fluctuation (BF).[18] The BE is formed by the Coulomb attraction between a trap centre and an electron-hole pair (i.e. exciton),[18] as schematically shown in Figure 2(a). According to Haynes rule,[18,32] an exciton that is bound to a neutral acceptor (i.e. A$^0$X) is (i) more stable than a free exciton and (ii) exhibits a higher $E_b^{BE}$ when it is associated with a trap that has a larger ionisation energy. Thus, to stabilise exciton states suitable for realising novel optoelectronic properties, such as perovskite RT STE, shallow defects need to be eliminated (to suppress charge screening), while deeper acceptor centres need to be introduced.

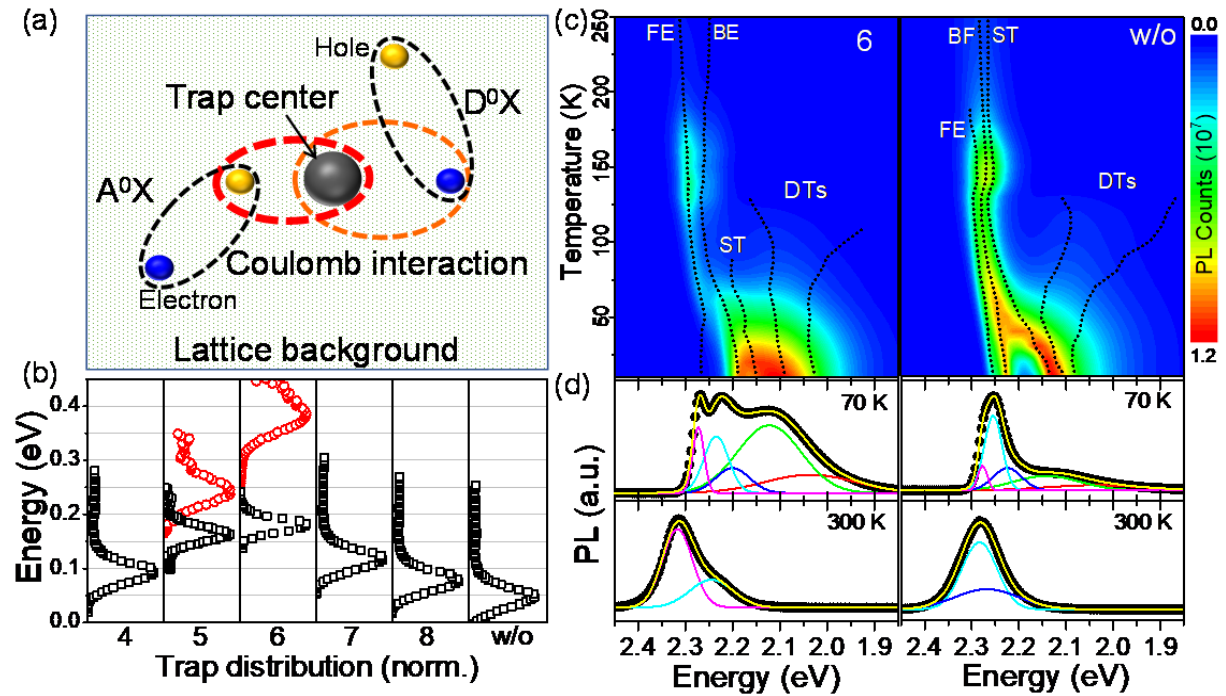

**Figure 2. Energy state engineering and exciton stabilisation of MAPbBr$_3$.** (a) Schematic diagram of the bound exciton complex, which is formed by the Coulomb attraction between a trap centre and an exciton. (b) Trap state distribution of the MAPbBr$_3$ film deposited by an anti-solvent method, where the time of chlorobenzene dropping is adjusted. (c) Temperature-dependent steady-state PL of the MAPbBr$_3$ with different deposition conditions (6 and w/o, excited at 450 nm, 5 nJ cm$^{-2}$). The emission peaks are systematically fitted and distinguished, as indicated by the black dashed lines (FE: free exciton, BE: bound exciton, ST: shallow trap, DTs: deep traps, BF: band fluctuation). For the 6-MAPbBr$_3$, FE and BE emission bands are observed across the entire temperature region. (d) Peaks fittings of emission spectra at 70K and 300K, respectively.

MAPbBr$_3$ is a ternary semiconductor, whose trap characteristics are largely dictated through self-doping mechanisms, e.g. vacancies and interstitials.[33] These can be modified through precisely tailored processing conditions. In this work, we have adopted an anti-solvent (chlorobenzene), solution-processed method to adjust the crystallization and growth dynamics of MAPbBr$_3$ thin films.[34] For simplicity, we have adopted a naming convention for the samples that relates the time at which the anti-solvent was cast onto the samples during spin-coating of the precursor solution (Supplementary Figure S2), i.e. the sample derived from chlorobenzene dropped after 6 seconds is denoted as 6-MAPbBr$_3$. Samples that did not have an anti-solvent dropped are denoted as w/o-MAPbBr$_3$. Variation in the processing conditions resulted in modified optical and structural characteristics, as verified optically, and through x-ray diffraction, light absorption and transient PL measurements (see Supplementary Figure S3-S7). These imply that the crystallization process has indeed been modified. Moreover, their trap state properties have been characterized using thermal admittance spectroscopy within an FTO/compact TiO$_2$/MAPbBr$_3$/Spiro-OMeTAD/Au device architecture at temperatures ranging from 150K to 330K (Supplementary Figure S8-S10).[30] All the samples possess an intermediate-high-frequency capacitance response, which arises from shallow trap states. However, the energy level of this shallow trap state is dependent on the sample (Figure 2(b)). For the w/o-MAPbBr$_3$, the trap state is located at ~0.05 eV above the valence band and possesses a broad Gaussian distribution of ~0.05 eV. Meanwhile, for 6-MAPbBr$_3$, this shallow defect is observed at ~0.18 eV and has a much narrower distribution of ~0.03 eV. These energy levels and their varying densities (Supplementary Figure 11) imply that the 6-MAPbBr$_3$ exhibits the smallest band-edge fluctuation. In addition to these shallow traps, the low-frequency capacitance response indicates that deeper trap states are also presented in the 5- and 6-MAPbBr$_3$ samples at ~0.25 eV and 0.38 eV, respectively. As discussed above, these deep traps are likely to be favourable towards stabilising BE states.

To further probe their impact on the photo emission properties of the samples, temperature-dependent photoluminescence (PL) measurements were carried out from 10K to 300 K. The results of the 6- and w/o-MAPbBr$_3$ samples are presented in Figure 2(c), with the additional samples presented in Supplementary Figure S12 and S13. To distinguish the fine structure of the energy states, the PL spectra across all the samples were de-convoluted (Figure 2(d), Supplementary Figure S14). The emission position, broadening and intensity of each contribution were mapped across the studied temperature range (Supplementary Figures S15-S17), which helps to distinguish the mechanism for each emission band. For each sample, up to five emission bands could be observed at temperatures less than 100K. Based on the above energy-state discussions, these have been assigned to FE, BE (or BF), ST and deep defects (DTs). As expected, the emission from DTs disappeared at temperatures higher than ~130K due to the orthorhombic-tetragonal phase transition.[31] The FE and BE emission positions were independent of the pump intensity, while the ST and DT emissions showed characteristic state-filling behaviour (Supplementary Figure S15). For the w/o-MAPbBr$_3$, its FE emission disappeared above 200K, while the BF and ST emission remained across the studied temperature range. This result shows that the BF emission is correlated to a high density of STs in the w/o-MAPbBr$_3$, which could screen the excitonic nature of this material. Meanwhile, the deeper ST and DT in the 6-MAPbBr$_3$ help to stabilise the exciton states. These observations are consistent with the above considerations of excitonic phenomena.

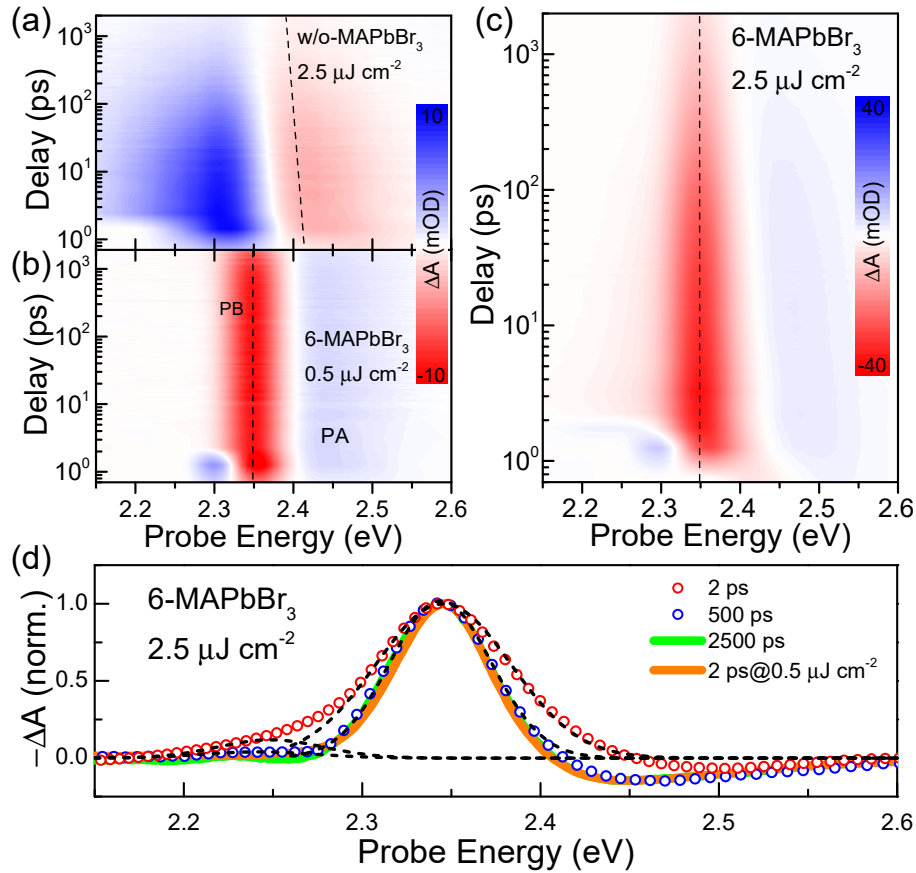

**Figure 3. Femtosecond transient absorption of the MAPbBr$_3$.** Two-dimensional pseudocolor plots ($\Delta A$) of transient absorption for the MAPbBr$_3$ films at 300K as functions of probe energy and time delay under different pump fluences (a. w/o: 2.5 μJ cm$^{-2}$, b. 6: 0.5 μJ cm$^{-2}$, (c) 6: 2.5 μJ cm$^{-2}$, pumped at 2.92 eV). The photobleaching (PB) peaks are illustrated by dashed black lines. (d) Photo-induced absorption spectra of the 6-MAPbBr$_3$ at different time delays and pump fluences. Gaussian-shaped dashed black lines are guides for the eye to illustrate the PB bands composed of FE and BE signals.

Femtosecond transient absorption (TA) has been used to further probe the temporal dynamics of the exciton states (Figure 3). When the MAPbBr$_3$ samples were pumped at 2.92 eV (425 nm), distinct photobleaching (PB) and photoinduced absorption (PA) bands could be observed. The PB peak of the w/o-MAPbBr$_3$ at ~2.4 eV exhibited a broad, high-energy distribution, with a progressive bathochromic-shift in time (Figure 3(a)). This should be a clear signature of carrier occupation and relaxation within the continuum band and bandedge states. Meanwhile, for the 6-MAPbBr$_3$, the spectral position of the main PB peak at ~2.35 eV was independent of the time delay and pump fluence (Figure 3(b) and (c)). These trends are characteristic signatures of excitonic processes,[11] which in this case can be ascribed to FEs. Together with the PL and TA, we can reasonably conclude that excitons and not free carriers dominate the emission properties of the 6-MAPbBr$_3$. We further note that for the 6-MAPbBr$_3$, a small energy difference of ~28 meV exists between the FE PB and PL peaks. This energy difference is similar to the longitudinal optical phonon (LO) energy of MAPbBr$_3$ (Supplementary Figure 16), implying an FE-LO process within the FE emission.

At a higher pump fluence of 2.5 μJ cm$^{-2}$, a PB contribution centred at ~2.25 eV becomes observable, consistent with the BE emission in the PL spectra. It has a rise time of ~1 ps and a decay lifetime of ~60 ps. This behaviour is consistent with an ultrafast exciton trapping process and subsequent fast spontaneous emission relaxation of the formed BE excited state. The BE state may simultaneously form when the FE relaxes within its Rydberg states. The bleaching characteristics of the identified BE state indicate that it possesses a much lower density-of-states than the FE contribution at ~2.35 eV. More detailed analysis on the high-energy PA[35] (Supplementary Figure 18-20) illustrate that no bandgap renormalization occurs in our studied pump fluence range. Therefore, we can confidently conclude that the FE and BE states of the MAPbBr$_3$ have been stabilised by energy state engineering.

**Perovskite excitonic STE**

To determine how the BE states influence the intrinsic STE behaviour of the MAPbBr$_3$, we have performed steady-state PL measurements (reflection configuration) with varying pump fluence under cavity free conditions. For the 6-MAPbBr$_3$, the FE and BE emission contributions can be observed across the entire studied fluence range without any observable change in the peak position. At ~2 μJ cm$^{-2}$, a much narrower emission peak centred at ~2.245 eV (552 nm) appears and becomes more prominent at higher fluences (Figure 4(a)). These are clear signatures of optically amplified spontaneous emission (ASE).[6] According to the pump-dependent emission intensity, the threshold fluence of the ASE for the 6-MAPbBr$_3$ is ~1.7 µJ cm$^{-2}$ (Figure 4(c) and Supplementary Figure 21). The corresponding threshold carrier density is calculated to be ~1.6×10$^{17}$ cm$^{-3}$. This value is almost one order of magnitude lower than previous results and the theoretical EHP threshold.[21] For reference, the 5-MAPbBr$_3$ also exhibited ASE with a similar spectral feature, albeit with a slightly higher threshold of ~2.4×10$^{17}$ cm$^{-3}$. As no signature of bandgap renormalization or excitonic-EHP transition was observed from the femtosecond TA, and the peak position overlays the BE emission, this optical amplification is attributed to BE STE. Notably, the ASE signals possessed no polarisation dependence, which is consistent with the samples being cavity-free and the excitons being randomly oriented within the polycrystalline film.

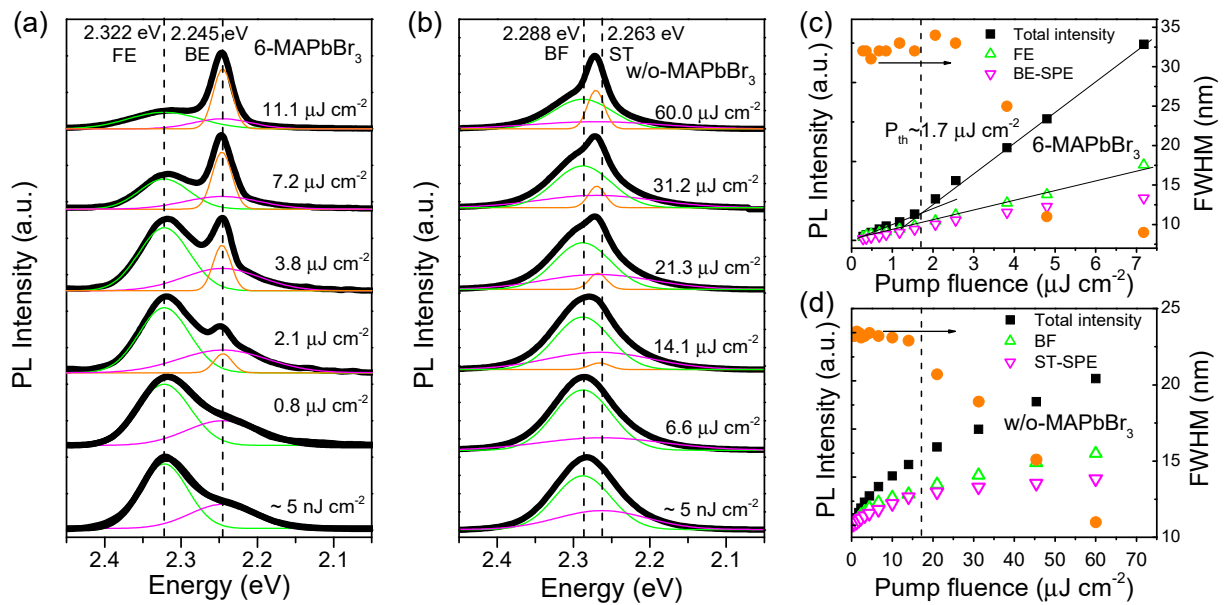

**Figure 4. Photo-pumped stimulated emission behaviour of MAPbBr$_3$ films.** (a and b) Time-integrated PL emission spectra of MAPbBr$_3$ films photo-pumped by 450 nm, 35 fs and 1 kHz pulses (a. 6-MAPbBr$_3$, b. w/o-MAPbBr$_3$). All the spectra are well fitted by considering the FE and BE spontaneous emission (BF and ST for the w/o-sample) and BE stimulated emission (ST for the w/o-sample). (c and d) FWHM and intensity of the emission peaks as a function of the pump fluence for (c) 6-MAPbBr$_3$ and (d) w/o-MAPbBr$_3$. The black dashed lines denote the threshold of stimulated emission. SPE: spontaneous emission.

Pump fluence dependent PL measurements of the w/o-MAPbBr$_3$ also demonstrated ASE. However, in this case, the band overlays the ST emission at 2.263 eV (548 nm) and possessed a much higher threshold fluence of ~15 μJ cm$^{-2}$ (Figure 4(b)). The peak position of this ASE is comparable to that previously reported,[8] indicating a similar EHP mechanism. The blue-shift of this ASE peak under higher pump fluences further supports this scenario. Due to the dominance of Auger recombination under these pumping conditions,[36] the optical amplification of the emission is weak (Figure 4(d)). Thus, although widely reported, the EHP STE has a greater reliance on the gain medium and cavity structure to fabricate laser performances with modest power consumption.

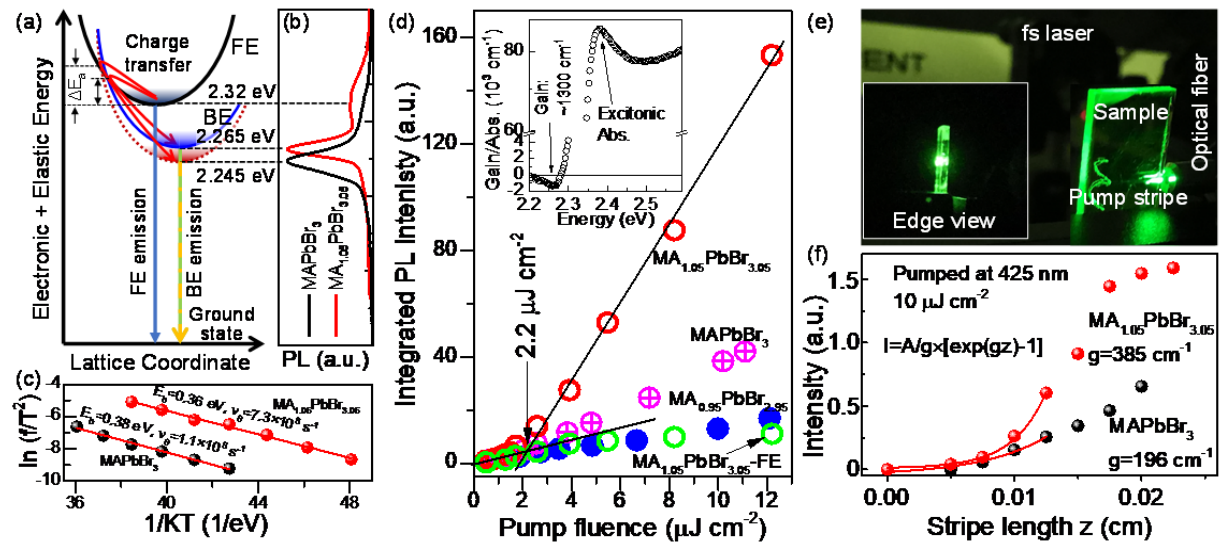

**Figure 5. Composition control to improve the charge transfer and the STE performance.** (a) Configuration coordinate diagram of total energy (electronic + elastic energies) of the perovskite to illustrate the charge transfer processes between the FE and BE states. Charge transfer occurs through the state coupling point between FE and BE to sustain energy conservation. The energy barrier ($\Delta E_a$) is the energy difference between the state coupling point and the bottom of the FE. Shifting of BE level in the energy axis results in a change in the $\Delta E_a$ and the charge transfer velocity. (b) STE spectra of the MAPbBr$_3$ and MA$_{1.05}$PbBr$_{3.05}$, which show a blue-shift of the BE energy level when the perovskite composition is adjusted from 1 to 1.05. (c) Arrhenius curves of deep trap states of the MAPbBr$_3$ and MA$_{1.05}$PbBr$_{3.05}$ samples to derive their attempt-to-escape frequency ($v_0$). (d) PL intensity of the MA$_{0.95}$PbBr$_{2.95}$, MAPbBr$_3$ and MA$_{1.05}$PbBr$_{3.05}$ as a function of pump fluence. The MA$_{1.05}$PbBr$_{3.05}$ shows an obvious enhancement in the BE STE and saturation in the FE emission under high pump fluences. The gain/absorption spectrum is included in the inset. (e)

Photo of the STE under an edge-coupled optical configuration, in which the film was pumped by an optical stripe and emission was detected from the film edge. (f) Stripe length (z) dependent emission intensity (solid circles: experimental data; solid lines: theoretical fittings to derive the gain coefficient ($g$)).

The low threshold STE within the 6-MAPbBr$_3$ is inherent to the BE optical gain mechanism; however, its super-radiance is not prominent from the pump-dependent PL curves. This arises because the emitted light only experiences a short gain length of up to ~200 nm within a reflection configuration and a large proportion of charges exist within the FE states. The linear relationship between the FE emission intensity and pump fluence indicate that the FE states are weakly correlated to the BE STE process. To overcome this intrinsic factor, enhancement in the charge transfer between the FE and BE states within the STE duration is required. This is a non-radiative, multi-phonon assisted process, whose rate is limited by an energy barrier ($\Delta E_a$, Figure 5(a)) that is determined by state coupling and energy conservation rules.[37-38] Based on these, at a given temperature, reducing the energy difference of the initial and final states (i.e. FE-BE) will promote the charge transfer.

Compared to a pure crystallization approach, element composition enables a more direct route to control the self-doping and thus the trap state distribution within the resulting perovskites.[39-40] As such, three samples with MABr/PbBr$_2$ ratios of 0.95, 1.0 and 1.05, respectively, were optimised. By using carrier density profiling measurement (Supplementary Figure 22), a much more concentrated carrier distribution was observed in the surface region of MA$_{0.95}$PbBr$_{2.95}$ film due to the MA or Br deficiency, while the MA$_{1.05}$PbBr$_{3.05}$ exhibited a similar carrier distribution to that of MAPbBr$_3$. The FE and BE emission bands were observed for all three samples, with the FE emission peaks all located at a similar energy position of ~2.32 eV. The high carrier density at the surface of the MA$_{0.95}$PbBr$_{2.95}$ led to a reduced STE performance, presumably due to charge screening effects. Meanwhile, the BE emission peak of the MA$_{1.05}$PbBr$_{3.05}$ was observed at a slightly higher energy of ~2.265 eV. This feature is also presented in the STE spectra shown in Figure 5(b), implying a non-negligible difference exists in the energy level of the BE state. To better understand the underlying traps associated with the BE states across these samples, we have performed temperature-dependent capacitance measurements to extract their energy level and attempt-to-escape frequencies ($v_0$).[30] As shown in Figure 5(c), the MA$_{1.05}$PbBr$_{3.05}$ exhibits a slightly lower trap energy compared to the stoichiometric sample, but with a 7-fold increase in $v_0$.

These modifications in the energy states are favourable for enhancing the charge transfer between FE and BE states within the sample, thus resulting in a more optimised three energy level system for achieving STE. The enhanced exciton transfer within the BE STE process of MA$_{1.05}$PbBr$_{3.05}$ is indicated by the femtosecond TA (Supplementary Figure 23-24), in which the FE PB shows a much faster decay in the early 10 ps and the initial depopulation of the BE PB is obviously slowed down. The impact is finally confirmed with a 3-fold enhancement in the optical amplification of the MA$_{1.05}$PbBr$_{3.05}$ sample (Figure 5(d)). Under an edge-coupled optical configuration, green STE emission can be clearly observed from the sample edge (Figure 5(e)). Notably, without a designed cavity for optical resonance, the STE light here did not show good directionality. A key material parameter that reflects this amplification process is the optical gain coefficient. By using a variable stripe length (VSL) method[8,28], the net gain coefficient of MAPbBr$_3$ and MA$_{1.05}$PbBr$_{3.05}$ was determined to be 196 cm$^{-1}$ and 385 cm$^{-1}$, respectively, under a pump density of 9×10$^{17}$ cm$^{-3}$. The gain spectrum of the MA$_{1.05}$PbBr$_{3.05}$ was further determined using pump-probe measurements taken at a 2 ps time delay, and a peak gain coefficient of up to ~1300 cm$^{-1}$ was derived (inset of Figure 5(d) and Supplementary Figure 25). Compared to VSL method, the pump-probe method minimizes the optical losses, which results in a more realistic measurement of a material's intrinsic gain.[41] The gain coefficients obtained here are among the highest for reported perovskites at such low pump density (Supplementary Table S3), and are even comparable to that of conventional quantum-well structures.[13, 41-42]

**Discussions**

In this work we have demonstrated room temperature STE arising from a bound exciton state for the first time within bulk perovskite MAPbBr$_3$ films through judicial energy state engineering. Its existence enables the material to possess a high gain coefficient of ~1300 cm$^{-1}$ and for optical amplification to be observed at a remarkably low threshold of ~1.6×10$^{17}$ cm$^{-3}$. These results highlight the tremendous promise that bulk metal halide perovskites offer towards laser application.

Until now, only several works have reported RT excitonic STE from bulk semiconductor films. Within inorganic CsPbCl$_3$ ($R_y^*$~64 meV) and CsPbBr$_3$ ($R_y^*$~35 meV) perovskite films, optical gain states were achieved at very high threshold powers of ~100 kW cm$^{-2}$.[43-44] These were considered to originate from inelastic exciton-exciton scattering. Meanwhile, exciton-cavity polaritons were reported within CsPbBr$_3$ and MAPbBr$_3$ perovskite nanowires, resulting in STE with reduced threshold fluences of ~15 μJ cm$^{-2}$.[45-46] For metal halides, such as CuBr and CuCl, which are well known to possess higher $R_y^*$ of >100 meV, RT excitonic STE has not been observed, even under very high pump powers.[47] RT STE has been reported for ZnO ($R_y^*$ ~60 meV) with a threshold power of 240 kW cm$^{-2}$, although the underlying mechanism is still under debate.[10, 19] These studies demonstrate that RT STE within bulk semiconductors has been achieved; however, due to a fundamental difference in the exciton emission mechanisms, they usually have much larger STE thresholds compared to what has been reported here. For quantum systems that have an enhanced exciton effect, such as CsPbBr$_3$ quantum dots,[25,28,48] comparable STE thresholds of several uJ cm$^{-2}$ have been demonstrated, but questions about their excitonic or EHP STE mechanisms remain unanswered.

The excellent STE characteristics of the studied hybrid perovskites arise from their distinct excitonic properties. Firstly, a $R_y^*$ of ~ 40 meV for MAPbBr$_3$ is higher than that of inorganic semiconductors with a similar $E_g$, such as ZnTe (~12 meV) and CdS (~25 meV), as well as wider $E_g$ candidates, such as GaN (~25 meV) and ZnS (~28 meV).[18] This large $R_y^*$ leads to a high Mott density, which helps to stabilise the RT excitonic phase at high carrier densities, enabling the population inversion of the BE state. Secondly, to our knowledge, an $E_b^{BE}$ of ~100 meV is among the largest reported for bulk semiconductors,[31] which has approached the upper limit of Haynes rule.[18]

It is evident that what enables such rich optoelectronic properties to emerge within perovskites is the underlying presence of tunable defects. The relatively wide and stable phase regime of this ternary semiconductor provides a large parameter space for self-doping and trap-state engineering using simple solution processes.[33] In this work we have focused on intrinsically derived modifications to the energy landscape. However, one can envisage the possibilities that further extrinsic dopants could provide, particularly by modifying their excitonic properties. With exciton states being sensitive to external fields, such as electric or magnetic, high sensitive electrical-optical and magnetic-optical coupled devices can be considered. Moreover, the spin properties of the BE state provides implications for quantum manipulation and communication.[49-50] The relatively large $R_y^*$ of the hybrid perovskite, and especially the stabilised bound exciton state, helps to make these applications practical at high temperatures, possibly even at room temperature. Each of these exciting prospects requires a greater level of understanding into the nature of the carrier and exciton states within perovskite materials.

**Acknowledgements**

This work was supported by Natural Science Foundation of China (No. 11874402, 91733301, 51761145042, 51421002, 51627803, 51872321) and the International Partnership Program of Chinese Academy of Sciences (No. 112111KYSB20170089). JJ would like to acknowledge funding through the ARC Centre of Excellence in Exciton Science (CE170100026).

**Author contributions**

J. Shi designed this study, and made experimental measurements, theoretical calculations and analysis; Y. Li prepared the sample and made photoluminescence measurements; J. Wu assisted in capacitance and transient absorption measurements and data extraction; H. Wu, Y. Luo and D. Li participated in organizing the measurement setups and assisted in the sample preparation; J. Jasieniak conceived and supported this project and participated in experimental and theoretical analysis; Q. Meng conceived and supported this project. All of authors contributed to the discussion. The manuscript was written by J. Shi, corrected by J. Jasieniak and Q. Meng and approved by all contributions.

**Additional information**

Supplementary information is available in the online version of the paper.

**Competing financial interests**

The authors declare no competing financial interests.

**Methods**

*Perovskite film deposition and characterisations.* The MAPbBr$_3$ and MAPbI$_3$ samples were deposited by using an anti-solvent method, where chlorobenzene is used as the anti-solvent. Briefly, a 1M MAPbBr$_3$ (or MAPbI$_3$) precursor solution in dimethylformamide (DMF) was firstly spread onto a quartz glass/compact Al$_2$O$_3$ substrate. A few of certain seconds after the beginning of spin coating at 4000 rpm, the chlorobenzene was dropping onto the film to produce a fast crystallisation of the perovskite. To adjust the film deposition process, the addition time for applying the anti-solvent was controlled as described in the main text. After 30 s of spinning, the film was transferred to a hot plate at 100 °C and annealed for 10 min. A 50 mg/ml PMMA solution was finally spin coated onto the perovskite film for protection. To adjust the composition, the precursors with different ratios of $MABr/PbBr_2$ were used. For each composition, the dropping time of anti-solvent has been optimised. The phase of the final perovskite was characterised by a Panalytical X-ray diffractometer. The averaged thickness of each film is measured by using a step profiler.

*Optical characterization.* The light transmission and reflectance of the perovskite film were measured with a UV-vis spectrometer (Shimadzu 3600). The light absorption is calculated from the transmission and reflectance. The steady-state and transient PL of the perovskite film were measured using an Edinburgh Fluorescence Spectrometer (FLS 920). A 445 nm pulsed diode laser (EPL-445, ~5 nJ $cm^{-2}$, 62 ps) was used as the excitation source. A circular adjustable neutral density filter was adopted to adjust the excitation intensity. The PL was collected in the reflectance mode, where a PMT together with a TCSPC module was applied to detect the time-resolved or time-integrated PL. For temperature dependent measurements, an ARS liquid helium cryostat was employed as the sample chamber under vacuum, where the temperature was controlled using a Lakeshore controller. The perovskite film on the glass substrate was directly fixed to the optical sample holder by screws. During the temperature adjustment, the system was firstly stabilised for at least 5 min at each temperature. Before the PL measurements, the perovskite sample was always kept in the dark without any bias light or laser illumination to avoid the possible production of charge accumulation and a local electric field.

For the STE measurement, a Coherent Astrella femtosecond amplifier (1 KHz, 35 fs, 7 mJ/pulse) together with an optical parameter amplifier (OPA, Opera-solo, 4 mJ/pulse split pulses for the OPA) was used as the pump source. The pump pulse was guided toward the sample vertically by planar mirrors and final a convex mirror to obtain a light spot diameter of ~12 mm. The laser power was measured by a coherent power meter. The averaged carrier density under photo-injection was calculated by considering the reflection and transmission. The emission from the sample was coupled into an optical fibre, placed at an angle relative to the incident laser of ~45°, through a long-pass filter and a convex lens. The spectrum was measured using an Avantes spectrometer (AvaSpec-ULS2048CL-EVO) with a CMOS detector operating at an integration time of 50 ms. The final spectrum were obtained by averaging 200 spectra with a total recording time of 10 s. For polarisation measurements, a linear polariser was added between the film and the optical fibre. For different polariser angles, the emission spectra were collected, and their intensity was compared directly. No difference in the STE intensity could be observed.

For the femtosecond transient absorption measurement, a high energy pulse generated from the femtosecond amplifier and the OPA were used as the pump light, and a white light generated by focusing the 800 nm (35 fs) pulse onto a 2 mm-thick sapphire and filtered by a short-pass filter (775 nm, Sigmakoki) was used as the probe light. The probe light was vertically focused onto the sample through a series of achromatic lens, with a final probe spot diameter of <1 mm. The pump light was focused onto the same centre by a concave mirror, with a spot diameter of ~5 mm. There was an angle of ~20° between the pump and probe light, that is, non-degenerate configuration. The time delay between these two pulses was controlled by using an Electric stage (Zolix) together with a hollow retroreflector (Newport). The transmitted probe light was detected by a CMOS optical fibre spectrometer (Avantes, AvaSpec-ULS2048CL-EVO) with a single-pulse mode (integrating time 30 μs). A pinhole 10 cm away from the sample is put in this light path to eliminate the light scattering from pump light and PL. With multiple spectra recording and calculating, the noise of the $\Delta A$ is lower than 0.1 mOD. The chirp of the probe pulse was calibrated by measuring the Kerr effect of $CS_2$. The VSL measurement was performed based on another CCD array optical fibre spectrometer (Ocean optics QE pro).

*Electrical characterisation.* The trap states of the $MAPbBr_3$ film were measured by using thermal admittance spectroscopy (TAS). The frequency-dependent capacitances of the FTO/compact $TiO_2$/$MAPbBr_3$/Spiro/Au cells were measured by using an electrochemical station (Princeton). The cell was placed inside a vacuum chamber of a low-temperature probe station (Lakeshore TTPX) with liquid nitrogen as the cooling source. The bandwidth of the electrical probe was 1 GHz. For capacitance measurement, the cell was always measured under dark conditions with no bias voltage being applied to the cell. The data analysis was performed according to published literature describing TAS. The trap state energy levels were derived from the Arrhenius curves. Mott-Schottky curves were also measured to help obtain trap density approximations.